\documentclass[%
 reprint,superscriptaddress,
 amsmath,amssymb,prb,
]{revtex4-2}

\usepackage{graphicx}
\usepackage{dcolumn}
\usepackage{bm}
\usepackage{color}
\usepackage{siunitx}
\usepackage{tabularx}
\usepackage{booktabs}
\usepackage{braket} 
\usepackage[version=3]{mhchem} 

\usepackage{xcolor}

\def \yc{\ce{YbBi2ClO4}}
\def \yi{\ce{YbBi2IO4}}
\def \yx{YbBi$_2$(Cl, I)O$_4$}
\def \degC{$^{\circ}$C~}
\def \TN{$T_{\mathrm N}$}

\def \qm{$\mathbf{q}_{\mathrm m}$}
\def \Jh{$J_{\mathrm{eff}}=1/2$}

\begin{document}

\preprint{APS/123-QED}

\title{Crystal field splittings and magnetic ground state of the square-lattice antiferromagnets \yc{} and \yi{} with \Jh{}} \thanks{This manuscript has been authored by UT-Battelle, LLC under Contract No. DE-AC05-00OR22725 with the U.S. Department of Energy.  The United States Government retains and the publisher, by accepting the article for publication, acknowledges that the United States Government retains a non-exclusive, paid-up, irrevocable, world-wide license to publish or reproduce the published form of this manuscript, or allow others to do so, for United States Government purposes.  The Department of Energy will provide public access to these results of federally sponsored research in accordance with the DOE Public Access Plan (http://energy.gov/downloads/doe-public-access-plan).}

\author{Pyeongjae Park}
\email{parkp@ornl.gov}
\affiliation{Materials Science \& Technology Division, Oak Ridge National Laboratory, Oak Ridge, Tennessee 37831, USA}

\author{Qianli Ma}
\affiliation{Neutron Scattering Division, Oak Ridge National Laboratory, Oak Ridge, Tennessee 37831, USA}

\author{G. Sala}
\affiliation{Oak Ridge National Laboratory, Oak Ridge, TN, 37831, USA}

\author{S. Calder}
\affiliation{Neutron Scattering Division, Oak Ridge National Laboratory, Oak Ridge, Tennessee 37831, USA}

\author{Douglas L. Abernathy}
\affiliation{Neutron Scattering Division, Oak Ridge National Laboratory, Oak Ridge, Tennessee 37831, USA}

\author{Matthew B. Stone}
\affiliation{Neutron Scattering Division, Oak Ridge National Laboratory, Oak Ridge, Tennessee 37831, USA}

\author{Andrew F. May}
\affiliation{Materials Science \& Technology Division, Oak Ridge National Laboratory, Oak Ridge, Tennessee 37831, USA}

\author{Andrew D. Christianson}
\email{christiansad@ornl.gov}
\affiliation{Materials Science \& Technology Division, Oak Ridge National Laboratory, Oak Ridge, Tennessee 37831, USA}


\begin{abstract}
We report on the crystal field level splitting and magnetic ground state of the \Jh{} square lattice antiferromagnets \yc{} and \yi{} using powder inelastic neutron scattering (INS) and neutron diffraction measurements. Both compounds exhibit a well-isolated $\Gamma_{7}$ doublet ground state under a tetragonal crystal field environment, confirming a robust \Jh{} picture with slight XY-type anisotropic character in the $g$-tensor. Notably, the ground state wave functions closely resemble the $\Gamma_{7}$ doublet expected in the perfect cubic limit, consistent with the nearly cubic ligand configuration of eight \ce{O^{2-}} ions surrounding Yb$^{3+}$. Below \TN{}\,=\,0.21\,K, \yi{} exhibits a stripe long-range magnetic order characterized by an ordering wave vector \qm{}\,=\,(1/2, 0, 0) or its symmetry-equivalent (0, 1/2, 0), with magnetic moments aligned along \qm{}. The ordered moment is approximately 79\% of the classical prediction, significantly larger than expected from the isotropic $J_{1}-J_{2}$ model, suggesting the possible involvement of exchange anisotropy in explaining this observation. We show that symmetry-allowed XXZ and bond-dependent anisotropic exchange terms in a square lattice can play a critical role in stabilizing the stripe order and suppressing the moment reduction as observed. These findings establish \yc{} and \yi{} as unique platforms for exploring rich \Jh{} magnetism from two less investigated perspectives: (i) on a square lattice and (ii) within a (nearly) cubic ligand environment.
\end{abstract}

\maketitle

\section{Introduction}
The $S = 1/2$ square lattice antiferromagnet (SqAF) serves as a textbook model in the study of low-dimensional quantum magnetism~\cite{Reger_1988, Dagotto_1989, QMC_1991, Schulz_1992,Chandra_1988}. The Heisenberg model with isotropic nearest-neighbor (NN) and second NN antiferromagnetic interactions (the $J_{1}$--$J_{2}$ model) has been extensively studied for its potential to manifest a quantum spin liquid state (QSL)~\cite{Balents_2012, Morita_2015, Gong_2014, Liu_2022_Tens}. The ratio $J_{2}/J_{1}$ indicates the degree of additional quantum fluctuations induced by exchange frustration, which are maximized at $J_{2}=0.5J_{1}$. Theoretical studies have shown that a QSL can emerge in the range $0.4<J_{2}/J_{1}<0.6$, separating regions of two well-known magnetic orders: N\'{e}el-type ($J_{2}/J_{1} < 0.4$) and stripe-type ($J_{2}/J_{1} > 0.6$)~\cite{Balents_2012, Morita_2015, Gong_2014, Liu_2022_Tens}. Notably, this phase diagram is very similar to that of the $J_{1}-J_{2}$ Heisenberg model for $S=1/2$ triangular lattice systems~\cite{TLAF_QSL1, TLAF_QSL2, TLAF_QSL3}.

A compelling extension to these isotropic quantum spin models is exploring the effect of exchange anisotropy on quantum fluctuations. Such anisotropies can involve Ising- or planar-type XXZ anisotropies~\cite{TLAF_aniso1, TLAF_Ising1, TLAF_Ising2, BLCTO}, and more intriguingly, symmetry-allowed bond-dependent anisotropy terms~\cite{honeycomb_kitaev, RuCl3_sci, kitaev_review, TLAF_aniso2, CoI2_nphys}, each of which capable of introducing unique magnetic properties. A key factor in the recent success of investigations on exchange anisotropy is the effective spin-1/2 degree of freedom in isolated Kramers doublet ground states (\Jh{}), where the intrinsic coupling between spin and orbital degrees of freedom enables significant exchange anisotropy~\cite{jackeli2009}. However, experimental studies of \Jh{} square lattice materials with exchange anisotropy have remained lacking, limiting insights into the anisotropic spin model in $S=1/2$ SqAFs.

\begin{figure}
\includegraphics[width=0.96\columnwidth]{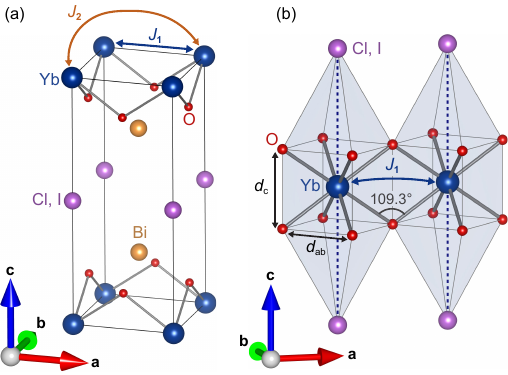} 
\caption{\label{crystal} Crystal structure of \yx{}. (a) The tetragonal unit cell of \yx{} [space group $P4/mmm$ (\#123)]. The Yb ions form a stack of square lattices well separated by Bi and Cl (I) layers. (b) The oxygen coordination of the Yb ions. The distances between Yb and O are all equal. The ratio of in-plane to out-of-plane O--O distances ($d_{\mathrm{ab}}/d_{\mathrm{c}}$) are 0.996 for both \yc{} and \yi{}~\cite{Park_YbBi2XO4}. This tiny elongation results in a Yb--O--Yb bond angle of $109.3^{\circ}$, nearly identical to the ideal tetrahedral angle of $\mathrm{cos}^{-1}({-1/3})=109.47^{\circ}$. Crystal structures were drawn with Vesta~\cite{vesta}.}
\end{figure}

\yi{} and \yc{} have recently been suggested as nearly ideal realizations of a 2D \Jh{} square lattice antiferromagnet~\cite{Park_YbBi2XO4, YbBi2IO4_PRB}. These compounds crystallize in a tetragonal structure ($P$4/$mmm$ space group), with Yb$^{3+}$ ions forming layered square lattices [Fig.~\ref{crystal}(a)]~\cite{Schmidt_2000_structure, Park_YbBi2XO4}. Each Yb$^{3+}$ ion is surrounded by a nearly cubic configuration of oxygen ligands with a marginal tetragonal distortion [$d_{ab}/d_{c}=0.996$, see Fig.~\ref{crystal}(b)]~\cite{Park_YbBi2XO4}. The \Jh{} nature of the Yb$^{3+}$ ions was inferred from heat capacity measurements, which revealed a total magnetic entropy of $R\ln(2)$~\cite{Park_YbBi2XO4,YbBi2IO4_PRB}. Temperature-dependent magnetic susceptibility [$\chi(T)$] of both compounds exhibits a broad maximum at a temperature $T_{\mathrm{max}}$ = 0.33 K (0.38 K) in \yi{} (\yc{}), indicating the quasi-2D nature of their magnetic Hamiltonian~\cite{Park_YbBi2XO4}. While Ref.~\cite{Park_YbBi2XO4} reported the onset of 3D long-range order below $T_N$ = 0.21\,K (0.25\,K) in \yi{} (\yc{}) based on a sharp lambda-like feature in $C(T)$, Ref.~\cite{YbBi2IO4_PRB} presented a qualitatively different $C(T)$ curve and suggested the absence of long-range order. Most importantly, the presence of a sizable $J_{2}$ was inferred from the much larger Curie-Weiss temperature magnitude ($|\theta_{\mathrm{CW}}|$) compared to $T_{\mathrm{max}}$, i.e., magnetic frustration,~\cite{Park_YbBi2XO4}. A quantitative estimation of $J_{2}/J_{1}$ using high-temperature series expansion analysis of the isotropic $J_{1}$--$J_{2}$ model, assuming $J_{1}>J_{2}$, resulted in $J_{2}/J_{1}$ close to 0.27 in these two compounds~\cite{Park_YbBi2XO4}, within the Ne\'el ordered phase.

These bulk properties strongly motivate further investigation into the microscopic magnetism of \yi{} and \yc{}, particularly regarding their spin Hamiltonians and associated quantum fluctuations. Notably, the simple theoretical analysis described above assumes an isotropic spin model, whereas the possibility of anisotropic exchange interactions between NNs should also be considered as potentially important in \Jh{} systems with edge-sharing ligand environments. Interestingly, \yi{} and \yc{} feature a cubic-like configuration of O$^{2-}$ ligands, entirely different from the octahedral ligand configurations that have been extensively studied~\cite{jackeli2009, Jackeli2010, honeycomb_kitaev, kitaev_review, Yb_aniso, motome_review, Pr_aniso}. This unique environment provides an opportunity to explore \Jh{} magnetism under entirely different conditions, potentially offering new insights into spin-orbital coupled interactions.

In this article, we present inelastic neutron scattering (INS) and neutron diffraction measurements of \yc{} and \yi{}, which reveal the crystal field splittings of Yb$^{3+}$ ions and the long-range order of the \Jh{} magnetic moments in \yi{}. The crystal field levels measured by INS confirm that Yb$^{3+}$ exhibits a Kramers doublet ground state separated by about 80\,meV from the first excited states in both \yc{} and \yi{}, supporting the picture of \Jh{} square-lattice magnetism at the magnetic energy scale ($\sim$1\,K). The crystal field Hamiltonian and the corresponding ground state wave function were derived by fitting the energy and dynamical structure factors of the measured excitations. This reveals a $\Gamma_{7}$ doublet ground state not far from the perfect cubic limit but exhibits slight XY-type anisotropy in their $g$-tensor. Powder neutron diffraction measurements of \yi{} at dilution refrigerator temperatures reveal magnetic Bragg peaks corresponding to an ordering wave vector \qm{} = (1/2, 0, 0) or its symmetry-equivalent directions, indicating a stripe magnetic order, antithetical to the proposed order based upon high temperature susceptibility measurements. Rietveld refinement shows that the magnetic moments point along \qm{}, with a magnitude 79\% of the classical prediction without quantum fluctuations. We discuss a spin model and corresponding degree of quantum fluctuations for \yi{} and \yc{} compatible with these findings, suggesting that these compounds may host sizable exchange anisotropy beyond the isotropic $J_{1}-J_{2}$ model.  

\section{Experimental details}
Polycrystalline \yc{} and \yi{} were prepared by solid-state reaction as described in Ref.~\cite{Park_YbBi2XO4}. \ce{LuBi2ClO4} and \ce{LuBi2IO4}, the nonmagnetic analogs of \ce{YbBi2ClO4} and \ce{YbBi2IO4}, were synthesized by similar methods using reactions at 950\degC for 100\,h followed by grinding and an additional heating step at 900\degC for 100\,h. The Lu-based compounds are used in the INS measurements to quantify the extent of the signal due to phonon scattering.

INS measurements of the crystal field splitting were performed with the ARCS spectrometer at the Spallation Neutron Source at the Oak Ridge National Laboratory (ORNL). Samples were loaded in aluminum (Al) sample cans with 1 atmosphere of helium gas. Data were collected using an incident energy of $E_i=$ 160\,meV, with the Fermi chopper set to 600\,Hz, providing an energy resolution of $=6.1$\,meV (full-width at half-maximum) at the elastic line. Background signals were estimated by measurements of nonmagnetic analogs \ce{LuBi2ClO4} and \ce{LuBi2IO4} under the same temperature, using identical Al sample cans. Empty can measurements were subtracted from all datasets unless otherwise noted. Data processing was carried out using the Dave~\cite{dave} and Mantid~\cite{mantid} software packages.

Powder neutron diffraction measurements of \yi{} were conducted with the HB-2A powder diffractometer at the High Flux Isotope Reactor at ORNL. To access temperatures below \TN{} = 0.21\,K, we used a dilution refrigerator. Approximately 9\,g of polycrystalline \yi{} were loaded into a standard copper sample can. The can was filled with over-pressurized helium gas ($p = 10$\,atm), which was necessary to ensure the sample fully thermalized at temperatures below \TN{}. Data were collected using a wavelength of 2.41\,\AA{} at temperatures of 0.04\,K, 0.25\,K, and 1.5\,K, with a counting time of approximately 8 hours per temperature. We used the FullProf software package to perform Rietveld refinement of crystal and magnetic structures~\cite{fullprof}.

\section{Results}
\subsection{Crystal field level splitting}
Yb$^{3+}$, a Kramers ion, has a $J=7/2$ multiplet ground state ($^{2}\mathrm{F}_{7/2}$ in term symbol formalism) that can split into a maximum of four Kramers doublets. In \yc{} and \yi{}, Yb$^{3+}$ ions lie in a tetragonal environment coordinated by eight oxygen ions and two I or Cl ions [Fig.~\ref{crystal}(b)], with the $4/mmm$ site symmetry. This configuration results in four Kramers doublets and therefore should have three excited levels.

\begin{figure}[h]
\includegraphics[width=1\columnwidth]{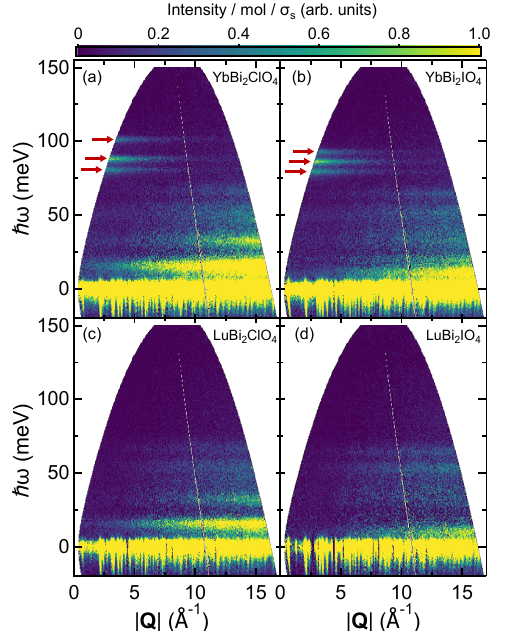} 
\caption{\label{CF_data} INS intensity measured at $T=10$\,K as a function of energy ($\hbar\omega$) and wave-vector transfer ($|\mathbf{Q}|$) for (a) \yc{}, (b) \yi{}, (c) \ce{LuBi2ClO4}, and (d) \ce{LuBi2IO4}. Data were acquired using the ARCS instrument as detailed in the text. The intensity has been normalized to the product of the number of measured moles of the sample and the total scattering cross-section per formula unit. Data have had an empty can background subtraction applied prior to normalization. All data are shown on the same intensity scale. Red arrows indicate the three excited crystal field levels of Yb$^{3+}$ in these materials.}
\end{figure}

Our powder INS measurements successfully identified these three excited crystal field levels in \yc{} and \yi{}. Figures \ref{CF_data}(a)--(b) show the powder INS spectra of \yc{} and \yi{} at 10\,K with an incident neutron energy of 160\,meV. Three clear, flat excitations are observed in both compounds as indicated by red arrows in Figs. \ref{CF_data}(a)--(b). They exhibit higher intensity in the low-momentum ($|\mathbf{Q}|$) region indicating the behavior of a magnetic form factor, and are absent in the INS spectra of the nonmagnetic analogs \ce{LuBi2ClO4} and \ce{LuBi2IO4}. These observations confirm that the excitation signals correspond to Yb$^{3+}$ crystal field levels.

Notably, all excited levels are located above 79\,meV in both compounds, indicating that the ground-state Kramers doublet is well isolated at temperatures comparable to the magnetic energy scale ($\sim1$\,K) of \yc{} and \yi{}. This ensures an exact \Jh{} picture, consistent with prior magnetic entropy measurements showing saturation at $R\mathrm{ln}(2)$~\cite{Park_YbBi2XO4}. In addition, no noticeable difference is observed between the spectra collected at 10\,K and 100\,K, as the activation energy ($\sim$80\,meV) still dominates over the thermal fluctuation energy at 100\,K.

The eigenvalues and relative dynamical structure factors of the three excited levels were analyzed quantitatively using constant wave-vector scans through the crystal field excitation spectra (integrated over the momentum range $4<|\mathbf{Q}|<5\,$\AA$^{-1}$), as shown in Figs.~\ref{CF_fit}(a) and \ref{CF_fit}(e). These values were obtained by simultaneously fitting the three peaks with Lorentzian profiles, in which we refined the amplitude (= structure factors), center position (= eigenvalues), linewidth, and a global constant background. A multiplicative scaling factor was applied to the overall intensity during the fit. 

{\renewcommand{\arraystretch}{1.1}
\begin{table}[!h]

\caption{Fitted eigenvalues ($\hbar\omega_{i}, i=1,2,3$) and relative spectral weights ($I_{i}/I_{j}$) of the three crystal field excitations in  for \yc{} and \yi{}.}
\begin{center}
\centering
\begin{tabularx}{\columnwidth}{X X X} 
\midrule \midrule
 & YbBi$_2$ClO$_4$ & YbBi$_2$IO$_4$ \\ 
\midrule
$\hbar\omega_{1}$\,(meV) & 80.78 & 79.61 \\ 
$\hbar\omega_{2}$\,(meV) & 88.12 & 86.32 \\ 
$\hbar\omega_{3}$\,(meV) & 101.31 & 92.65 \\ 
$I_{1}/I_{2}$ & 0.560 & 0.756 \\
$I_{3}/I_{2}$ & 0.591 & 0.469  \\
\midrule \midrule
\end{tabularx}
\end{center}
\label{tab_CFfit}
\end{table}

The resultant eigenvalues and structure factor ratios, summarized in Table~\ref{tab_CFfit}, provide crucial information of the tetragonal crystal field Hamiltonian and corresponding eigenstates of each compound. With the $c$-axis as the unique tetragonal axis [see Fig.~\ref{crystal}(b)], the crystal field Hamiltonian ($\mathcal{\hat H}_{\mathrm{CF}}$) can be expressed as~\cite{abragam2012,CF_tetra}:

\begin{eqnarray}
\mathcal{\hat H}_{\mathrm{CF}} &=& B_{2}^{0}\mathcal{\hat O}_{2}^{0} + B_{4}^{0}\mathcal{\hat O}_{4}^{0} + B_{4}^{4}\mathcal{\hat O}_{4}^{4} + B_{6}^{0}\mathcal{\hat O}_{6}^{0} + B_{6}^{4}\mathcal{\hat O}_{6}^{4},
\label{CFHam}
\end{eqnarray}

\noindent where $\mathcal{\hat O}_{n}^{m}$ are conventional Stevens operators~\cite{stevens}, and $B_{n}^{m}$ are the corresponding crystal field parameters. This Hamiltonian yields the following four doublet states represented by two 2D irreducible representations, $\Gamma^{t}_{7}$ and $\Gamma^{t}_{6}$~\cite{CF_tetra}:

\begin{eqnarray}
\ket{\Gamma^{\mathrm{t}}_{7,1}} &= \pm c_{1} \ket{\pm\frac{5}{2}} \pm c_{2} \ket{\mp \frac{3}{2}}, \nonumber \\
\ket{\Gamma^{\mathrm{t}}_{7,2}} &= \mp c_{1} \ket{\pm\frac{5}{2}} \pm c_{2} \ket{\mp \frac{3}{2}}, \nonumber\\
\ket{\Gamma^{\mathrm{t}}_{6,1}} &= \pm a_{1} \ket{\mp\frac{7}{2}} \pm a_{2} \ket{\pm \frac{1}{2}}, \nonumber\\
\ket{\Gamma^{\mathrm{t}}_{6,2}} &= \mp a_{1} \ket{\mp\frac{7}{2}} \pm a_{2} \ket{\pm \frac{1}{2}}, \\\nonumber
\label{Eq:doublets}
\end{eqnarray}

\noindent where $M_{J}$ in $\ket{M_{J}}\equiv\ket{J=7/2,M_{J}}$ denotes the eigenvalue of $\hat{J_{z}}$, and the normalized coefficients $c_{1,2}$ and $a_{1,2}$ depend on $B_{n}^{m}$.

\begin{figure}[t]
\includegraphics[width=1\columnwidth]{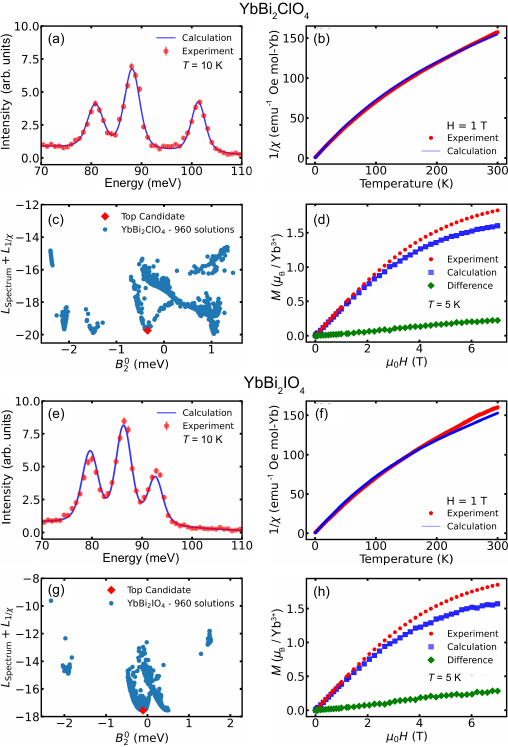} 
\caption{\label{CF_fit} Quantitative analysis of the crystal field excitations and magnetic susceptibility. (a) Energy dependence of the measured INS intensity ($4$\,\AA$^{-1}<|\mathbf{Q}|<5$\,\AA$^{-1}$) at 10\,K (red circles) from \yc{}. The solid blue line represents the calculated INS cross-section based on the best-fitted crystal field parameters $B^{n}_{m}$ from TABLE.~\ref{tab_Hamfit}. The line spans from and to the energy transfer range included in the fit. (b) Measured and calculated inverse magnetic susceptibility for \yc{}, with the latter derived from the model in TABLE.~\ref{tab_Hamfit}. (c) Five-dimensional goodness-of-fit landscape for the \yc{} Hamiltonian, projected onto the $B^{0}_{2}$ axis. The red diamond marks the solution presented in TABLE.~\ref{tab_Hamfit}. (d) Comparison of the measured and calculated field-dependent magnetization at 5\,K. (e)--(h) Same as (a)--(d), but for \yi{}. Simulations in (e)--(h) are based on the solution with $B^{2}_{0}=-0.107$\,meV (see TABLE~\ref{tab_Hamfit}.}
\end{figure}

While determining $B_{n}^{m}$ is typically achieved by minimizing the $\chi^{2}$ metric between the observed and calculated crystal field spectra, this approach could potentially end up with multiple possible solutions. To ensure physically meaningful results based on sufficient constraints, we simultaneously fit the crystal field spectrum and the temperature-dependent magnetic susceptibility. This was performed by minimizing a custom-defined loss function using CrysFieldExplorer \cite{MaCEF}:

\begin{eqnarray}
L_{\mathrm{tot}} &=& L_{\mathrm{Spectrum}} + L_{\mathrm{1/\chi}},
\label{Eq:Loss}
\end{eqnarray}

\noindent where $L_{\mathrm{Spectrum}}$ and $L_{\mathrm{1/\chi}}$ are the loss functions concerning the crystal field spectra [Fig.~\ref{CF_fit}(a) and \ref{CF_fit}(e)] and the inverse-susceptibility [Fig.~\ref{CF_fit}(b) and ~\ref{CF_fit}(f)], respectively. $L_{\mathrm{Spectrum}}$ is further defined as $L_E + L_\mathrm{Intensity}$. Their explicit functional expressions are:

\begin{eqnarray}
&L_E = \log_{10}\left(\sum_i^n det\{(E_{\mathrm{exp}}[i]+E_{\mathrm{cal}}[0])\mathrm{I} -H_{\mathrm{CF}} \}^2\right), \\
&L_{\mathrm{Intensity}} = \dfrac{\sqrt{\sum_i^n ({I_{\mathrm{true}}[i]-I_{\mathrm{calc}}}[i])^2}}{\sqrt{(\sum{I_{\mathrm{true}}}[i])^2}}, \\
&L_{1/\mathrm{\chi}} = \dfrac{\sqrt{({1/\mathrm{\chi}_{\mathrm{exp}}-1/\mathrm{\chi}_{\mathrm{calc}}})^2}}{\sqrt{({1/\mathrm{\chi}_{\mathrm{exp}}})^2}}.
\label{Eq:Loss}
\end{eqnarray}

where the summation runs through 1 to $n$-th crystal field levels of Yb$^{3+}$. $E_\mathrm{exp}[i]$ represents the energy levels of the observed crystal field excitations determined by fitting Lorentzian convolved Gaussians to the constant wave-vector scans in Figs.~\ref{CF_fit}(a) and ~\ref{CF_fit}(e). $E_\mathrm{calc}[0]$ is eigenvalues obtained from the diagonalization of the crystal field Hamiltonian $H_{\mathrm{CF}}$ (Eq.~\eqref{CFHam}). $\mathrm{I}$ is the identity matrix. $I_{\mathrm{true}}[i]$ and $I_{\mathrm{calc}}[i]$ are the normalized observed and calculated intensities for the $i$-th CEF excitation. $\mathrm{\chi}_{\mathrm{exp}}$ and $\mathrm{\chi}_{\mathrm{calc}}$ denotes the experimentally measured and calculated magnetic susceptibility data. Details of the construction of the loss functions are demonstrated in Ref.~\cite{MaCEF}.

The results of minimizing $L_{\mathrm{tot}}$ throughout the five-dimensional parameter space are shown in Fig.~\ref{CF_fit}(c) and Fig.~\ref{CF_fit}(g). The key term in Eq.~\eqref{CFHam} concerning the tetragonal nature of \yc{} and \yi{} is $B_{2}^{0}\mathcal{\hat O}_{2}^{0}$. The other four terms are also allowed under cubic symmetry. Given the importance of $B_{2}^{0}$, we minimized $L_{\mathrm{tot}}$ for a fixed $B_{2}^{0}$ value, repeating this process across a range of $-2.5\,\mathrm{meV}<B_{2}^{0}<1.5\,\mathrm{meV}$.

For \yc{}, good fits were achieved by three solutions at $B_{2}^{0}=-1.48, -0.344,$ and $1.06$\,meV, as shown in Fig.~\ref{CF_fit}(c) as a deep local minimum of the $L_{\mathrm{tot}}$ landscape. However, a careful analysis of their resultant ground states reveals that the cases of $B_{2}^{0}=-1.48$ and $1.06\,$meV exhibit unphysical features inconsistent with bulk characterization results, which is described in Appendix A. For \yi{}, we found two potential solutions at $B_{2}^{0}=-0.107$ and $0.487$\,meV [Fig.~\ref{CF_fit}(g)], but further analysis could not conclusively distinguish between these solutions, as described below. 

{\renewcommand{\arraystretch}{1.1}
\begin{table}[!t]

\caption{The optimized Stevens operator coefficients ($B_{n}^{m}$) for \yc{} and \yi{}, and corresponding wave function coefficients and $g$-factors. The uncertainty in the fitted $B_{n}^{m}$ values is approximately 4\,\% of their value; see the main text. For \yi{}, two viable solutions are presented, both of which exhibit nearly identical ground state wave functions, making them equivalent from the \Jh{} magnetism perspective. In all cases, the four doublets in Eq.~\eqref{Eq:doublets} are arranged in ascending energy order as $\Gamma^{t}_{7,1}-\Gamma^{t}_{6,1}-\Gamma^{t}_{7,2}-\Gamma^{t}_{6,2}$ (also see Appendix A). The $d_{\mathrm{ab}}/d_{\mathrm{c}}$ values are adapted from Ref.~\cite{Park_YbBi2XO4}.}
\begin{center}
\centering
\begin{tabularx}{\columnwidth}{X X X X} 
\midrule \midrule
 & YbBi$_2$ClO$_4$ & YbBi$_2$IO$_4$ & YbBi$_2$IO$_4$\\ 
\midrule
$B_{2}^{0}$\,(meV) & -0.3441 & -0.1068 & 0.4870 \\ 
$B_{4}^{0}$\,(meV) & 0.0496 & 0.0433 & 0.0421 \\ 
$B_{4}^{4}$\,(meV) & 0.2740 & 0.2862 & 0.2780 \\ 
$B_{6}^{0}$\,(meV) & -0.0003 & -0.0004 & 0.0001\\ 
$B_{6}^{4}$\,(meV) & -0.018 & -0.0168 & -0.0179 \\
\midrule
$c_{1}$ & -0.8143 & -0.7851 & -0.7868 \\ 
$c_{2}$ & 0.5804 & 0.6194 & 0.6172 \\ 
$a_{1}$ & -0.9999 & -0.9820 & -0.052 \\ 
$a_{2}$ & 0.0080 & 0.1890 & -0.9986 \\ 
\midrule
Ground state & $\ket{\Gamma^{\mathrm{t}}_{7,1}}$ & $\ket{\Gamma^{\mathrm{t}}_{7,1}}$ & $\ket{\Gamma^{\mathrm{t}}_{7,1}}$\\
\midrule

$g_{\mathrm{ab}}$ & 3.74 & 3.85 & 3.85 \\ 
$g_{\mathrm{c}}$ & 2.63 & 2.21 & 2.23 \\
\midrule
$d_{\mathrm{ab}}/d_{\mathrm{c}}$ & 0.996 & 0.996 & 0.996 \\  
\midrule \midrule
\end{tabularx}
\end{center}
\label{tab_Hamfit}
\end{table}

The full parameter set for each solution and the corresponding wave functions are summarized in TABLE.~\ref{tab_Hamfit}, along with their coefficients $c_{1,2}$ and $a_{1,2}$. The uncertainty in the fitted $B^{n}_{m}$ values is estimated to be approximately 4\,\% of their magnitude, as such variations did not produce any noticeable visual disparity in our fit results (Fig.~\ref{CF_fit}). Notably, except for $B^{0}_{2}$, the fitted $B^{n}_{m}$ values are very similar across all solutions, remaining nearly within the aforementioned uncertainty range. Indeed, the corresponding ground state doublet of each solution ($\Gamma^{t}_{7,1}$) is close to each other, with coefficients $c_{1}$ and $c_{2}$ differing only slightly. This consistency aligns with previous findings that \yi{} and \yc{} show nearly the same magnetic properties~\cite{Park_YbBi2XO4}. The corresponding in-plane ($g_{ab}$) and out-of-plane ($g_{c}$) $g$-factors of the ground state doublet, also listed in TABLE.~\ref{tab_Hamfit}, show a consistent profile across all solutions: $g_{ab}>g_{c}$, confirming an XY-like anisotropic character of the single-ion Yb$^{3+}$ magnetism in both \yi{} and \yc{}.

Another noteworthy observation is that the two excited doublets, $\ket{\Gamma^{t}_{6}}$, are nearly pure $\ket{\pm\frac{1}{2}}$ or $\ket{\pm\frac{7}{2}}$ states, with minimal superposition between them. The key difference between the two solutions for \yi{} in TABLE.~\ref{tab_Hamfit} lies in the energy hierarchy of $\ket{\pm\frac{1}{2}}$ and $\ket{\pm\frac{7}{2}}$. Each solution assigns the observed excitations at $\hbar\omega_{1}$ and $\hbar\omega_{3}$ as either $\ket{\frac{1}{2}}$ and $\ket{\frac{7}{2}}$ or $\ket{\frac{7}{2}}$ and $\ket{\frac{1}{2}}$. While the data at hand cannot resolve which assignment is correct, we emphasize that both solutions yield identical magnetic properties due to their quantitatively identical ground state wave functions. The first excited level--either $\ket{\pm\frac{1}{2}}$ or $\ket{\pm\frac{7}{2}}$--is significantly separated from the ground state and would not affect the magnetic properties at temperatures comparable to the magnetic energy scale ($\sim1$\,K).

Of further interest is the comparison between the observed $M-H$ curves in the paramagnetic regime ($T\gg|\theta_{\mathrm{CW}}|$) and the predictions from the fitted ground state doublet [Fig.~\ref{CF_fit}(d) and \ref{CF_fit}(h)]. While the experimental curves are qualitatively consistent with theoretical expectations, showing saturation around $1.67\mu_{\mathrm{B}}/\mathrm{Yb}^{3+}$ (with an averaged $g$-factor of $(2g_{ab}+g_{c})/3\sim3.34$ for both compounds), we found apparent underestimation of magnetization values by the calculation in the high-field regime. The origin of this discrepancy remains unclear, but notably the deviation scales linearly with field strength.

Finally, it is useful to compare our results to the case of perfect cubic symmetry, characterized by the constraints $B^{0}_{2}=0$, $B^4_{4}=5B^{0}_{4}$, and $B^4_{6}=-21B^{0}_{6}$. In the limit $|B^{0}_{4}| \gg |B^{0}_{6}|$ (see TABLE.~\ref{tab_Hamfit}), this symmetry yields the following $\Gamma_{7}$ doublet ground state~\cite{cubic_CEF, CF_tetra}:

\begin{eqnarray}
\ket{\Gamma^{\mathrm{c}}_{7}} &= \frac{\sqrt{3}}{2} \ket{\pm\frac{5}{2}} - \frac{1}{2}\ket{\mp \frac{3}{2}}.
\label{Eq:Gcubic}
\end{eqnarray}

This state is not far from the ground states found in \yi{} and \yc{}, with the fitted $c_{1}$ and $c_{2}$ coefficients showing a slight deviation from the cubic limit. Thus, despite the presence of the tetragonal characteristics, such as anisotropic $g$-factors, to some extent \yi{} and \yc{} would retain the essence of \Jh{} magnetism typical of perfect cubic ligand environments. This aspect might be relevant to the nearly cubic arrangement of the eight oxygen anions surrounding Yb$^{3+}$ [Fig.~\ref{crystal}(b)], as evident in the near-unity ratio of in-plane to out-of-plane O--O distances (see TABLE~\ref{tab_CFfit})~\cite{Park_YbBi2XO4}.

Thus, \yi{} and \yc{} offer a rare platform for studying \Jh{} magnetism in an edge-sharing (nearly) cubic environment. We note that, the magnitude of $B^{0}_{2}$ in TABLE~\ref{tab_Hamfit}, which is the largest or second largest among the coefficients, should not be misinterpreted as it is the dominant term in the Hamiltonian. This is because, the contribution of higher order terms in Eq.~\ref{CFHam} accompanies a few order-of-magnitude larger multiplication factor to $B^{n}_{m}$ in the solution and thus actually dominates over the lower order terms when their $B^{n}_{m}$ coefficients are comparable~\cite{CF_tetra}. This becomes evident when examining the analytic form of, for instance, the phase factor $\phi_{7}$, which is directly related to the wave function through $c_{1}=\cos{(\phi_{7}/2)}$~\cite{CF_tetra}:

\begin{eqnarray}
\tan{\phi_{7}} = {-\frac{2}{\sqrt{3}} \left(\frac{-15B^{4}_{4}+105B^{4}_{6}} {B^{0}_{2}-50B^{0}_{4}-1470B^{0}_{6}}\right)}.
\label{Eq:phi7}
\end{eqnarray}

We note that the definition of $B^{n}_{m}$ used in this work [Eq.~\eqref{CFHam}] is different from that in Ref.~\cite{CF_tetra} by constant multiplication factors. 

\begin{figure}[!th]
\includegraphics[width=0.99\columnwidth]{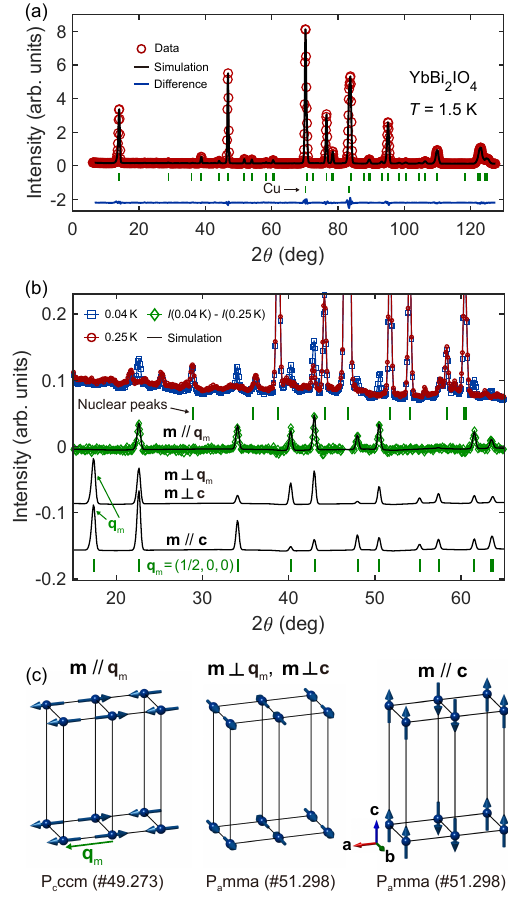} 
\caption{\label{NPD} Powder neutron diffraction profiles of \yc{} and \yi{} collected at temperatures below and above \TN{}. (a) Powder diffraction pattern of \yi{} at $T$ = 1.5\,K ($T>$\,\TN{}). (b) Magnetic diffraction profile of \yi{} extracted from the difference between the 0.04\,K ($T<T_{\mathrm{N}}$, blue-colored) and 0.25 K ($T>T_{\mathrm{N}}$, red-colored) data; see green data points. Black solid lines are the three simulated diffraction patterns for magnetic moments aligned along \qm{}, perpendicular to both \qm{} and the $c$-axis, and along the $c$ axis, respectively. (c) Three stripe magnetic structure models [with \qm{} = (1/2, 0, 0)], with magnetic moments pointing along \qm{}, perpendicular to both \qm{} and the $c$-axis, and along the $c$-axis, respectively. The magnetic space groups (using Belov–Neronova–Smirnova notation) for each illustration are noted below.}
\end{figure}

\subsection{Magnetic structure}
The magnetic long-range order in \yi{} was investigated using powder neutron diffraction with a dilution refrigerator. The crystal structure of \yi{} has previously been studied via Rietveld refinement of time-of-flight neutron diffraction profiles collected at $T$ = 30\,K~\cite{Park_YbBi2XO4}. We conducted the same analysis on new data collected at $T = 1.5 K$, still far above \TN{} [Fig. \ref{NPD}(a)]. The refinement result is nearly identical to that reported in our previous work~\cite{Park_YbBi2XO4}, with no significant differences observed.

Neutron diffraction measurements at 40 mK, well below \TN{}, revealed the emergence of magnetic Bragg peaks in \yi{} [Fig. \ref{NPD}(b)], demonstrating the onset of long-range order. This diffraction profile corresponds to a magnetic ordering wave vector \qm{} = (1/2, 0, 0) (reciprocal lattice units, r.l.u.) or its symmetry-equivalent \qm{} = (0, 1/2, 0). For clarity, we will describe our findings based on \qm{} = (1/2, 0, 0) hereafter. This ordering wave vector generates an alternating configuration of ferromagnetic spin chains, typically referred to as a stripe~\cite{Balents_2012} or columnar-type magnetic order~\cite{Mustonen_2018}. Notably, in the isotropic $S=1/2$ $J_{1}$--$J_{2}$ model, such a magnetic structure is expected only for $J_{2}/J_{1}>0.6$, which is significantly larger than the $J_{2}/J_{1}$ ratio predicted for \yi{} from high-temperature series expansion analysis~\cite{Park_YbBi2XO4}. Additionally, the parallel alignment between 2D square-lattice layers implies the sign of interlayer interactions: assuming additional interlayer couplings are negligible, the NN interlayer interaction is inferred to be ferromagnetic.

The direction of magnetic moments was unambiguously determined by the absence of intensity at the ($\frac{1}{2}$00) at $2\theta=18^{\circ}$ [Fig. \ref{NPD}(b)]. This indicates that magnetic moments are parallel to \qm{} [i.e., along the a-axis for \qm{} = (1/2, 0, 0)], as only this orientation suppresses the ($\frac{1}{2}$00) peak by making its neutron polarization factor zero. This is evident in the comparison between our data to simulated diffraction patterns for the \qm{} = (1/2, 0, 0) magnetic structure with moments along the $a$, $b$, and $c$ directions [black solid lines in Fig.~\ref{NPD}(b)]. Also, the observed magnetic Bragg peak intensities are well described by the model with moments parallel to \qm{} (the $a$-axis). Finally, quantitative refinement of the peak intensities yields an ordered moment magnitude of 1.53(4)$\mu_{\mathrm{B}}/\mathrm{Yb}^{3+}$ for \yi{}.

\begin{figure}[th]
\includegraphics[width=0.9\columnwidth]{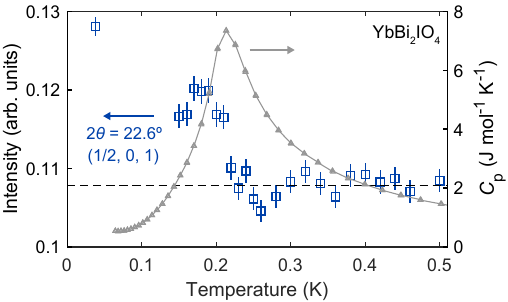} 
\caption{\label{Tdep} Temperature dependence of the (1/2, 0, 1) magnetic Bragg peak in \yi{}. For comparison, we overplot the temperature-dependent heat capacity of \yi{}~\cite{Park_YbBi2XO4}.}
\end{figure}

Fig.~\ref{Tdep} shows the temperature dependence of the diffraction intensity at $2\theta=22.6^{\circ}$, corresponding to the ($\frac{1}{2}$01) peak. The data reveal the emergence of the magnetic Bragg peak below around 0.21\,K. This is consistent with \TN{}\,=\,0.21\,K for \yi{} obtained from heat capacity measurements~\cite{Park_YbBi2XO4}, also shown in Fig.~\ref{Tdep}.

\section{Discussion}

The combination of crystal field splitting and the magnetic ground state presented in this work provides crucial insights concerning the nature of magnetism in \yc{} and \yi{}. The observed stripe magnetic order highlights the importance of exchange interactions beyond antiferromagnetic $J_{1}$, as the $J_{1}$-only model predicts a Ne\'el-type order with \qm{}=(1/2, 1/2, 0). Given the similarity in bulk properties~\cite{Park_YbBi2XO4} and ground state wave functions, we infer that \yc{} likely adopts the same \qm{}=(1/2, 0, 0) ground state, although the moment direction could differ. The simplest interpretation for this observation is the presence of strong antiferromagnetic $J_{2}$: in the isotropic $J_{1}-J_{2}$ model, $J_{2}/J_{1}>0.6$ stabilizes the stripe order~\cite{Balents_2012, Morita_2015, Gong_2014,Liu_2022_Tens}. 

However, the isotropic $S=1/2$ $J_{1}$--$J_{2}$ model reveals limitations in describing other observations. Combining neutron diffraction (1.53(4)$\mu_{\mathrm{B}}/\mathrm{Yb}^{3+}$) with crystal field splitting analysis ($g_{ab}=3.85$), the observed ordered moment magnitude in \yi{} corresponds to the 79(2)\,\% of the classical prediction. This 21\,\% reduction is much smaller than the 40\,\% reduction typically expected from the isotropic $S=1/2$ $J_{1}$--$J_{2}$ model (in the absence of significant exchange frustration, i.e., $J_{2}\gg J_{1}$ or $J_{1} \gg J_{2}$), as confirmed by both experiment and theories~\cite{A⁢MoOP⁢O4⁢Cl, Li2VOSiO4, park_ZVPO, Balents_2012, SqAF_XXZ}. Moreover, the reduction increases further as $J_{2}$ and $J_{1}$ become more comparable since it drives the model closer to the QSL region, $0.4<J_{2}/J_{1}<0.6$. While interlayer coupling ($J_{c}$) can remedy this discrepancy, a large separation between adjacent square lattice layers [Fig.~\ref{crystal}(a)], a broad maximum in temperature-dependent magnetic susceptibility~\cite{Park_YbBi2XO4,YbBi2IO4_PRB}, and preliminary estimations based on density functional theory electronic structure~\cite{YbBi2IO4_PRB} consistently suggest its very weak strength. Thus, incorporating $J_{c}$--likely only a few percent of $J_{1}$ or $J_{2}$--may not suppress the moment reduction enough to the level observed~\cite{SqAF_inter,park_ZVPO}.

The above conclusion motivates us to explore the possibility of magnetic anisotropy in \yi{} and \yc{}. The symmetry-allowed interaction matrix between NN Yb$^{3+}$ sites can be expressed as follows (using the parametrization notation of $\Delta$ and $J_{\pm\pm}$ commonly applied in other lattice models~\cite{TLAF_aniso2, ross2011}):

\begin{align}
\hat{\mathcal{H}}_{ij} =
\mathbf{S}_i^T
\begin{pmatrix}
J_{1} + 2J_{\pm\pm}\tilde{c}_{ij} & 0 & 0 \\
0 & J_{1} - 2J_{\pm\pm}\tilde{c}_{ij} & 0 \\
0 & 0 & \Delta J_{1}
\end{pmatrix}
\mathbf{S}_j,
\label{Eq:matrix}
\end{align}

where $\mathbf{S}_{i}$ is the (effective) spin operator acting on the localized moment at site $i$ of a square lattice, $\tilde{c}_{ij}=\cos(2\phi_{ij})$, with $\phi_{ij}$ being the angle between the a-axis and the bond vector connecting sites $i$ and $j$. In other words, $\tilde{c}_{ij}=\pm1$ for the bonds along a- and b-directions, respectively.

\begin{figure}[t]
\includegraphics[width=1\columnwidth]{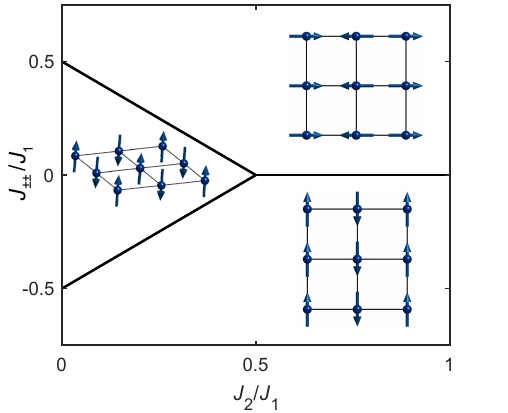} 
\caption{\label{pdg} Classical phase diagram of the $J_{1}-J_{2}-J_{\pm\pm}$ model obtained by classical Monte Carlo simulations combined with simulated annealing. For $J_{\pm\pm}=0$, spin vectors in each phase are not fixed to the direction illustrated in the figure.}
\end{figure}

The parameter $\Delta$ quantifies the degree of XXZ exchange anisotropy. The in-plane spin configuration found in \yi{} indicates that only the easy-plane type ($\Delta<1$) is compatible. While reducing $\Delta$ from 1 does not significantly alter the phase boundaries between Ne\'el-type and stripe magnetic orders in the isotropic $J_{1}-J_{2}$ model, it gradually suppresses quantum fluctuations inherent to the isotropic model~\cite{SqAF_XXZ}. Notably, this behavior is very similar to the role of $\Delta<1$ in $S=1/2$ triangular lattice systems~\cite{BLCTO,TLAF_aniso1,TLAF_aniso2}. Calculations in Ref.~\cite{SqAF_XXZ} suggest that achieving a moment reduction as small as 20\,\% roughly requires $\Delta<0.7$, a significant level of easy-plane anisotropy. Thus, strong easy-plane XXZ anisotropy could be one possibility of the suppressed moment reduction.

Another component, $J_{\pm\pm}$, represents bond-dependent exchange anisotropy. Notably, $J_{\pm\pm}=\pm0.5J_{1}$ corresponds to the pure quantum compass model~\cite{SqAF_compass}. The classical phase diagram of the $J_{1}-J_{2}-J_{\pm\pm}$ model provides qualitative insights into the influence of $J_{\pm\pm}$ on the competition between Ne\'el-type and stripe long-range orders (Fig.~\ref{pdg}). Interestingly, $J_{\pm\pm}$ favors the observed stripe order over the N\'{e}el order, regardless of its sign, significantly expanding the stripe phase compared to the isotropic $J_{1}-J_{2}$ model. We found that $J_{\pm\pm}>0$ produces a stripe configuration consistent with our observations, where the magnetic moments align parallel to \qm{} (Fig.~\ref{pdg}). This makes our findings in \yi{} compatible with either a large $J_{2}/J_{1}>0$, a large $J_{\pm\pm}/J_{1}>0$, or a combination of both.

More importantly, increasing $|J_{\pm\pm}|$ is expected to gradually suppress the moment reduction in the stripe phase. In this phase, $J_{\pm\pm}$ induces an energy gap in the Goldstone mode at $\mathbf{Q}=\mathbf{q}_{\mathrm{m}}$, reducing zero-point fluctuations and thereby suppressing the moment reduction. A quantitative estimation of this effect requires advanced quantum-mechanical simulations of the $J_{1}-J_{2}-J_{\pm\pm}$ model. Such simulations are particularly valuable for examining how the known range of the QSL phase in the isotropic model ($0.4<J_{2}/J_{1}<0.6$) evolves under finite $J_{\pm\pm}$.

Given the compatibility of our observations with the two symmetry-allowed anisotropies, assessing their feasibility from a microscopic structural perspective is a crucial direction for further investigation. Numerous studies have shown that \Jh{} moments in edge-shared octahedral ligand configurations can exhibit substantial anisotropic exchange interactions, in both transition metal and rare-earth based systems~\cite{jackeli2009, Jackeli2010, honeycomb_kitaev, kitaev_review, Yb_aniso, motome_review, Pr_aniso}. Since \yi{} and \yc{} also feature \Jh{} moments in an edge-shared ligand environment, there is no \textit{a priori} reason to exclude the possibility of such interactions. However, the interaction profiles strongly depend on the specific orbital arrangements of the magnetic ions and ligands. Unfortunately, theoretical studies on the cubic ligand network is still lacking. In Appendix B, we note both the important similarities and differences between octahedral and cubic ligand configurations, highlighting the challenges in drawing intuitive conclusions for cubic environments.

Experimental insights into \Jh{} magnetism in cubic ligand environments are also limited due to the scarcity of relevant materials other than \yi{} and \yc{}. A comparable material class is the Yb-based pyrochlore oxides Yb$_{2}M_{2}$O$_{7}$ ($M$ = Ti, Ge, Sn), where Yb$^{3+}$ is surrounded by an edge-shared polyhedron of eight oxygens~\cite{Yb_pyrochlore}. Notably, INS studies on these materials revealed significant exchange anisotropy~\cite{227_aniso1,227_aniso2, 227_aniso3}. That being said, their oxygen configurations are substantially distorted from the ideal cubic form, limiting the direct applicability of these findings to \yi{} and \yc{}.

Overall, our experimental results suggest the potential presence of exchange anisotropy in \yi{} (and possibly \yc{}). Determining the generalized spin Hamiltonian of these systems using cold-neutron spectroscopy is essential to unambiguously confirm this possibility. Regardless, our findings suggest that \yi{} and \yc{} open a new avenue for exploring the rich \Jh{} magnetism from two perspectives: (i) on a square lattice (ii) within a nearly cubic ligand environment--both of which have been far less studied compared to the extensively investigated triangular or honeycomb systems with octahedron environments. 
\vspace{\baselineskip}

\section{Conclusion}
We have studied the crystal field level splitting and magnetic ground state of the \Jh{} square lattice antiferromagnets \yc{} and \yi{}. Under the tetragonal crystal field environment, both compounds exhibit a well-isolated $\Gamma^{t}_{7}$ doublet ground state, confirming a robust \Jh{} picture with a slight XY-type anisotropic g-tensor ($g_{ab}/g_{c}$\,=\,1.42 and 1.73 in \yc{} and \yi{}, respectively). The refined ground state wave functions in both compounds resemble the $\Gamma^{c}_{7}$ doublet expected in the perfect cubic limit, which might be attributed to marginal tetragonal distortion~\cite{Park_YbBi2XO4} in the nearly cubic ligand configuration of eight O$^{2-}$ ions. \yi{} exhibits a stripe long-range order below \TN{}\,=\,0.21\,K, indicating the presence of interactions beyond the isotropic $J_{1}$. The refined ordered moment is $\sim$79\,\% of its classical prediction, considerably larger than that predicted by the isotropic $S=1/2$ $J_{1}-J_{2}$ model ($\sim$60\,\%), which would not be fully attributed to interlayer coupling. Interestingly, symmetry-allowed XXZ and bond-dependent anisotropy terms both favor the observed magnetic structure and the suppressed moment reduction. Further investigations, including theoretical estimations of the exchange interaction profile and experimental studies using cold-neutron spectroscopy, are essential to validate the suggested anisotropic spin Hamiltonian. Our findings suggest that \yi{} and \yc{} provide a rare opportunity to explore rich \Jh{} magnetism (i) on a square lattice and (ii) within a nearly cubic ligand environment.

\begin{acknowledgments}
 This research was supported by the U.S. Department of Energy, Office of Science, Basic Energy Sciences, Materials Science and Engineering Division. This research used resources at the High-Flux Isotope Reactor and Spallation Neutron Source, a DOE Office of Science User Facility operated by the Oak Ridge National Laboratory. The beam time was allocated to ARCS on proposal number IPTS-31742.1 and HB-2A on proposal number IPTS-31409.1.
\end{acknowledgments}

\providecommand{\noopsort}[1]{}\providecommand{\singleletter}[1]{#1}%

\appendix

\clearpage
\section{Assessing other potential solutions of the crystal field excitation analysis}
Tables~\ref{tab:yc_Eig} and \ref{tab:yi_Eig} present the eigenvalues and eigenstates corresponding to the optimal crystal field parameters listed in Table~\ref{tab_Hamfit}, specifically for the case of $B^{2}_{0}=-0.1068\,$meV for \yi{}. While the fitting result indicates a good agreement from a few other parameter sets for \yc{}, these alternative solution candidates, with $B_{2}^{0}=-1.48$ and $1.06\,$meV [see Fig.~\ref{CF_fit}(c)], do not accurately represent the magnetic properties of \yc{}. The complete parameter sets of these alternative solutions (in meV) are as follows:

\begin{eqnarray}
&B^{2}_{0}=1.06,\, B^{4}_{0} = -0.050,\, B^{4}_{4} = -0.154,\nonumber \\
&B^{6}_{0} = 0.0026,\, B^{6}_{4}=-0.0131,
\end{eqnarray}

\begin{eqnarray}
&B^{2}_{0}=-1.48,\, B^{4}_{0} = -0.043,\, B^{4}_{4} = 0.174,\nonumber  \\
&B^{6}_{0} = -0.0007,\, B^{6}_{4}=0.0214.
\end{eqnarray}

Notably, both solutions yield a $\Gamma^{t}_{6}$ doublet ground state, different from those given by the solutions presented in TABLE~\ref{tab_Hamfit}. The corresponding $g$-factors are:

\begin{eqnarray}
g_{ab}=4.085,\,\,g_{c}=0.170,
\end{eqnarray}

\begin{eqnarray}
g_{ab}=0.907,\,\,g_{c}=6.185.
\end{eqnarray}

However, the average $g$-factors for these solutions, calculated as $g_{\mathrm{avg}}=(2g_{ab}+g_{c})/3$, are 2.780 and 2.666, respectively. These values are significantly smaller than the $g_{\mathrm{avg}}$ value for \yc{} obtained from Curie-Weiss fitting and saturated magnetization of the $M-H$ curve~\cite{Park_YbBi2XO4, YbBi2IO4_PRB}. On the other hand, the chosen solution in TABLE~\ref{tab_Hamfit} giving a $\Gamma^{t}_{7}$ ground state yields a correct $g_{\mathrm{avg}}$.

\section{Comparing octahedral and cubic ligand configurations}
Here, we highlight some crucial similarities and differences between \Jh{} magnetism in the widely studied traditional octahedral ligand configuration and our case with a nearly cubic environment~\cite{Yb_aniso}. First, the doublet ground state found in \yi{} and \yc{} (Table.~\ref{tab_Hamfit}), which is not far from the perfect cubic case [Eq.~\eqref{Eq:Gcubic}], falls into the same $\Gamma_{7}$ representation as the \Jh{} doublet realized in transition metal systems with octahedral ligand configurations (e.g., $d^{5}$ iridates and ruthenates, and $d^{7}$ cobaltates)~\cite{Yb_aniso, motome_review}. It is also noteworthy that 4$f^{1}$ rare-earth systems with the octahedral environment (e.g., Ce$^{3+}$), which were suggested to exhibit strong antiferromagnetic Kitaev interactions, also configures the same $\Gamma_{7}$ ground state doublet as $\Gamma^{t}_{7}$ in Eq.~\eqref{Eq:doublets}~\cite{motome_review}. Second, concerning the NN superexchange interactions, both configurations consist of two symmetry-equivalent exchange paths [see Fig.~\ref{crystal}(b)]. On the other hand, a contrasting feature is the bond angle: unlike a 90$^{\circ}$ bond angle found in the octahedral configuration, the cubic configuration is characterized by a tetrahedral bond angle, $\mathrm{cos}^{-1}({-1/3})=109.47^{\circ}$. Each aspect crucially influences the underlying hopping processes and the net exchange interaction matrix, thereby difficult to infer what would happen in the cubic case without explicit calculation of the hopping processes.

\renewcommand{\arraystretch}{1.1}
\begin{table*}
\caption{Eigenvalues and Eigenvectors of the Yb$^{3+}$ crystal field levels in \yc{}}
\begin{ruledtabular}
\begin{tabular}{c|cccccccc}
E (meV) &$| -\frac{7}{2}\rangle$ & $| -\frac{5}{2}\rangle$ & $| -\frac{3}{2}\rangle$ & $| -\frac{1}{2}\rangle$ & $| \frac{1}{2}\rangle$ & $| \frac{3}{2}\rangle$ & $| \frac{5}{2}\rangle$ & $| \frac{7}{2}\rangle$ \tabularnewline
 \hline
0.000 & 0.0 & 0.0 & 0.5804 & 0.0 & 0.0 & 0.0 & -0.8143 & 0.0 \tabularnewline
0.000 & 0.0 & 0.8143 & 0.0 & 0.0 & 0.0 & -0.5804 & 0.0 & 0.0 \tabularnewline
80.78 & -0.9999 & 0.0 & 0.0 & 0.0 & 0.0080 & 0.0 & 0.0 & 0.0 \tabularnewline
80.78 & 0.0 & 0.0 & 0.0 & 0.0080 & 0.0 & 0.0 & 0.0 & -0.9999 \tabularnewline
88.12 & 0.0 & -0.5804 & 0.0 & 0.0 & 0.0 & -0.8143 & 0.0 & 0.0 \tabularnewline
88.12 & 0.0 & 0.0 & -0.8143 & 0.0 & 0.0 & 0.0 & -0.5804 & 0.0 \tabularnewline
101.31 & 0.0 & 0.0 & 0.0 & 0.9999 & 0.0 & 0.0 & 0.0 & 0.0080 \tabularnewline
101.31 & -0.0080 & 0.0 & 0.0 & 0.0 & -0.9999 & 0.0 & 0.0 & 0.0 \tabularnewline
\end{tabular}\end{ruledtabular}
\label{tab:yc_Eig}
\end{table*}

\begin{table*}
\caption{Eigenvalues and Eigenvectors of the Yb$^{3+}$ crystal field levels in \yi{}}
\begin{ruledtabular}
\begin{tabular}{c|cccccccc}
E (meV) &$| -\frac{7}{2}\rangle$ & $| -\frac{5}{2}\rangle$ & $| -\frac{3}{2}\rangle$ & $| -\frac{1}{2}\rangle$ & $| \frac{1}{2}\rangle$ & $| \frac{3}{2}\rangle$ & $| \frac{5}{2}\rangle$ & $| \frac{7}{2}\rangle$ \tabularnewline
 \hline
0.000 & 0.0 & 0.0 & 0.6194 & 0.0 & 0.0 & 0.0 & -0.7851 & 0.0 \tabularnewline
0.000 & 0.0 & 0.7851 & 0.0 & 0.0 & 0.0 & -0.6194 & 0.0 & 0.0 \tabularnewline
79.61 & -0.9820 & 0.0 & 0.0 & 0.0 & 0.1890 & 0.0 & 0.0 & 0.0 \tabularnewline
79.61 & 0.0 & 0.0 & 0.0 & 0.1890 & 0.0 & 0.0 & 0.0 & -0.9820 \tabularnewline
86.32 & 0.0 & -0.6194 & 0.0 & 0.0 & 0.0 & -0.7851 & 0.0 & 0.0 \tabularnewline
86.32 & 0.0 & 0.0 & -0.7851 & 0.0 & 0.0 & 0.0 & -0.6194 & 0.0 \tabularnewline
92.65 & 0.0 & 0.0 & 0.0 & 0.9820 & 0.0 & 0.0 & 0.0 & 0.1890 \tabularnewline
92.65 & -0.1890 & 0.0 & 0.0 & 0.0 & -0.9820 & 0.0 & 0.0 & 0.0 \tabularnewline
\end{tabular}\end{ruledtabular}
\label{tab:yi_Eig}
\end{table*}

\end{document}